\documentclass[prl,twocolumn,showpacs,superscriptaddress]{revtex4-1}
\usepackage{graphicx}
\usepackage[usenames,dvipsnames]{xcolor}
\usepackage{epstopdf}
\usepackage{amssymb}
\usepackage{amsmath,amsthm,amssymb}
\usepackage{latexsym}
\usepackage{bm}
\usepackage{dcolumn}
\usepackage{braket}
\usepackage[normalem]{ulem}
\usepackage{times}

\newcommand{\SCO}{$\rm Sr_2CuO_3$}

\begin{document}

\title{
Magnons in \SCO: possible evidence for Goldstone-Higgs interaction
in a weakly ordered spin-1/2 chain antiferromagnet. }

\author{E.~G.~Sergeicheva}
\affiliation{P.~Kapitza Institute for Physical Problems, 119334, Moscow, Russia}

\author{S.~S.~Sosin}
\email{sosin@kapitza.ras.ru}
\affiliation{P.~Kapitza Institute for Physical Problems, 119334, Moscow, Russia}

\author{L.~A.~Prozorova}
\affiliation{P.~Kapitza Institute for Physical Problems, 119334, Moscow, Russia}

\author{G.~D.~Gu}
\affiliation{CMPMSD, Brookhaven National Laboratory, Upton, New York 11973, USA}

\author{I.~A.~Zaliznyak}
\email{zaliznyak@bnl.gov}
\affiliation{CMPMSD, Brookhaven National Laboratory, Upton, New York 11973, USA} \email{zaliznyak@bnl.gov}

\date{\today}

\begin{abstract}
We report on an electron spin resonance (ESR) study of a nearly one-dimensional (1D) spin$-1/2$ chain system, \SCO, with extremely weak symmetry-breaking order. The ESR spectra at $T > T_N$, in the disordered Luttinger-spin-liquid phase
reveal ideal Heisenberg-chain behavior with only very small, field-independent linewidth, $\sim 1/T$. In the ordered state, below $T_N$, we identify antiferromagnetic resonance (AFMR) modes,
which are well described by pseudo-Goldstone magnons in the model of a collinear biaxial antiferromagnet.
Additionally, we observe a major resonant response of a special nature, which we attribute to magnon interaction with the Higgs amplitude (longitudinal) mode in a weakly ordered antiferromagnet.
\end{abstract}

\maketitle

The symmetry broken states and quasi-particle excitations in condensed matter explore much of the same physics as field theories of particles in the Universe, while presenting an advantage of being precisely tunable and accessible in a laboratory-scale experiments. While experimental studies of Higgs particles or quark confinement in hadrons require giant colliders, such as LHC, or RHIC, important insights into the underlying physics can be obtained by studying model material systems such as quantum magnets, topological insulators, and ferroelectric and disordered superconductor systems \cite{Merchant_etal_Ruegg_NatPhys2014,Sato_etal_Takahashi_NatPhys2011,Lin_etal_Cheong_NatPhys2014,Sherman_etal_Dressel_NatPhys2015}. Transition metal compounds with weakly interacting Heisenberg spin-1/2 chains present particularly favorable opportunity for exploring these concepts. 

The ground state of an isolated chain is disordered, with de-confined fractional spin-1/2 excitations (spinons) forming continuum of physically accessible spin-1(2,3,...) states \cite{Walters_etal_Zaliznyak_NatPhys2009,Lake_etal_NatPhys2010,Zaliznyak_NatMat2005}. This state, known as Luttinger liquid (LL), is quantum-critical, so that even a tiny inter-chain coupling in a real material leads to a three-dimensional (3D) magnetic ordering at $T < T_N$.
Spontaneous symmetry breaking by magnetic order imposes linear attractive potential, which at low energy confines pairs of spin-1/2 spinons into spin-1 magnons, in accord with the Goldstone theorem which requires gap(mass)less excitations \cite{GoldstoneSalamWeinberg_PhysRev1952}. In the case
of a collinear antiferromagnetic (AFM) order, these are well-known transverse spin waves \cite{Andreev_JETP1978,AndreevMarchenko_1980}. The interaction of such Goldstone magnons with the Higgs \cite{Higgs_RMP2013,PekkerVarma_AdvCMP2015} amplitude mode of the order parameter is usually discarded while it grows in importance near the quantum critical point (QCP), where the symmetry-breaking 
is weak, and its amplitude fluctuations are significant. What is the energy scale where this interaction occurs, and what is the spectral weight involved -- are coherent magnons detectable at all near the QCP, when the symmetry breaking is weak? What are the roles of the amplitude fluctuations of the order parameter (Higgs mode), and of the longitudinal magnon mode predicted by the chain-mean-field theory \cite{Schulz_PRL1996,EsslerTsvelikDelfino_PRB1997}, the evidence for whose existence obtained in neutron experiments is still controversial \cite{Lake_NatMater2005,Lake_PRB2005,Zheludev_PRL2002}?

Motivated by these questions, we carried out ESR experiments aimed at a comparative study of magnetic resonance in the ordered and LL phases of a chain cuprate, \SCO, at frequencies of the microwave field, $\nu$, probing magnetic excitations with energies $h\nu \lesssim k_B T_N$. Surprisingly, not only we have been able to identify Goldstone magnons, modes corresponding to the transverse oscillations of the magnetic order parameter and a hallmark of a spinon confinement, but we have also discovered an unusual excitation with strongly field-dependent energy gap (mass). This mode is not predicted by the low-energy hydrodynamic theory of spin dynamics, which instead predicts a field-independent pseudo Goldstone spin wave \cite{Andreev_JETP1978,AndreevMarchenko_1980}. Therefore, the observed massive mode can only arise from the short-wavelength behavior, such as the interaction of the Goldstone magnon with the Higgs mode.

\SCO\ is a unique example of a nearly 1D $S=1/2$ chain antiferromagnet. It has a body-centered orthorhombic crystal structure (space group $Immm$) composed of chains of corner-sharing CuO$_4$ square plaquettes in the $(ab)$-plane, running along the $b$-axis of the crystal. The strong Cu-O hybridization results in an extremely strong Cu-O-Cu in-chain
superexchange, $J\approx 2800$~K, \cite{Suzuura_PRL1996,Walters_etal_Zaliznyak_NatPhys2009,Motoyama_PRL1996}. Small orbital overlaps between the planar CuO$_4$ plaquettes on neighbor chains yield a much smaller inter-chain coupling, $J^{\prime}/J \lesssim 5\cdot 10^{-4}$, resulting in an almost ideal spin-chain structure. Hence, \SCO\ undergoes a phase transition into an antiferromagnetically ordered state only below the N\'{e}el temperature $T_N = 5.5(1)$~K~$\approx2\cdot10^{-3}J$, in a very close proximity of the 1D LL quantum-critical state.
Strong quantum fluctuations result in an ordered moment of only $\langle\mu\rangle = 0.06\mu_{\rm B}$, as was determined from neutron scattering and $\mu$SR experiments \cite{Kojima_PRL1997}. Consequently, \SCO\ presents an ideal model material for exploring effects of an extremely weak symmetry breaking in a system of coupled quantum-critical spin-1/2 chains with fractional spinon excitations, the emergence of Goldstone and Higgs modes resulting from spinon confinement, and the corresponding dimensional cross-over regimes.

\begin{figure}[!t]
\centerline{\includegraphics[width=.8\columnwidth]{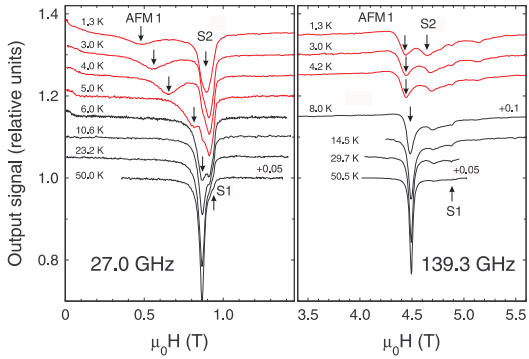}}
\caption{Temperature dependence of magnetic resonance spectra observed at $\nu = 27.0$~GHz (left panel) and $\nu = 139.3$~GHz (right panel) for magnetic field $H \parallel c$-axis of the sample. The signals are normalized to a unit level at maximum, and consecutively shifted by +0.05, from bottom to top. Red and black curves correspond to ordered and spin-liquid phases, respectively, and arrows mark different resonance peaks discussed in the text.} \label{temp_c}
\vspace{-0.25in}
\end{figure}
The ESR experiments were carried out on a high-quality single crystal sample ($m \approx 0.056$~g), similar to the ones used in our previous studies \cite{Walters_etal_Zaliznyak_NatPhys2009}. The sample was oriented using the tabletop Laue Xray. The magnetic resonance spectra were examined using a set of home made transmission-type microwave spectrometers with cylindrical and rectangular cavities covering the frequency range $22-140$~GHz. A magnetic field up to 12~T was supplied by the superconducting magnet. The temperature of the experiment varied from 0.5~K (with the $^3$He cryostat insert) to 50~K. For the complementary elastic neutron scattering measurements, a larger piece of a similar \SCO\ crystal was placed in a $^4$He flow cryostat on SPINS spectrometer at NIST Center for Neutron Research; the fixed scattered neutron energy of 3.7~meV, BeO filter after sample, and beam collimations $37'-80'-80'-240'$ from sample to detector were used.

Typical resonance absorption spectra of \SCO\ recorded at low and high frequencies of the microwave field are presented in Fig.~\ref{temp_c}. At $T>T_N$, the spectrum consists of an intense principal line and two weak satellites (marked S1 and S2). An excellent fit to the field profile of the ESR absorption signal for all measurement frequencies is obtained by using Lorentzian profile for the principal line and Gaussians for the satellites, resulting in three fitting parameters for each resonance line: the signal amplitude, $A$, the resonance field, $H_{res}$, and the half width at half maximum (HWHM) linewidth, $\Delta H$. 
At $T>T_N$ all observed resonance modes have linear frequency-field dependence typical of a paramagnet, $h\nu =g\mu_B H_{res}$ ($h$ is the Planck constant, $\mu_B$ is the Bohr magneton), with the $g$-factor values depending on the direction of the applied field. For the main line, the $g$-tensor components in the principal crystal axes are, $g^a=g^b=2.03\pm 0.02$, $g^c=2.22\pm 0.02$, consistent with the $(ab)$-plane geometry of the Cu $d_{x^2-y^2}$ orbital in \SCO\ \cite{Walters_etal_Zaliznyak_NatPhys2009}. For the satellite peak S1, $g_1^a=g_1^b=2.22\pm 0.02$, $g_1^c=2.03\pm 0.02$, while the S2 mode has isotropic $g$-factor, $g_2=2.11\pm 0.02$. As explained in detail in \cite{Supplementary}, these resonance lines correspond to two types of paramagnetic defects whose relative concentrations can be evaluated from the corresponding integral signal intensities normalized to the main signal: $n_{1} \approx 2 \cdot 10^{-5}$, $n_{2} \approx 1 \cdot 10^{-3}$, thus, establishing the exceptional quality of our single crystal samples \cite{Kojima_etal_PRB2004,Sirker_etal_PRL2007,Supplementary}. In the following, we therefore focus the attention at the properties of the main spin system.


\begin{figure}[!t]
\centerline{\includegraphics[width=0.9\columnwidth]{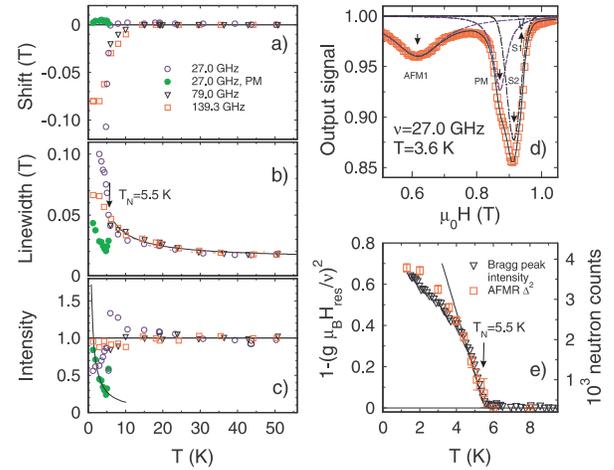}}
\caption{The resonance field shift (a), the line width (HWHM) (b), and the integral intensity (c) of the principal ESR line as a function of temperature, measured at three excitation frequencies: 27.0~GHz ({\Large $\circ$} and {\Large $\bullet$} are the two spectral components below $T_N$), 79.0~GHz ($\triangle$), and 139.3~GHz ($\square$), for  $H\parallel c$. Solid lines in panels (b), (c) show $\alpha +\beta /T$ and $C/T$ fits, respectively, for the width of the main line and the intensity of the residual paramagnetic component associated with defects. (d) The decomposition of a typical low-frequency resonance line in the ordered phase into Lorentzian (AFM1, PM) and Gaussian (S1, S2) spectral components. (e) The square of the AFMR gap, $\Delta^2_1(T)/\nu^2$, obtained from the temperature dependence of the resonance field measured at $\nu = 27.0$~GHz; the intensity of the $(0, 0.5, 0.5)$ magnetic Bragg peak vs temperature from neutron diffraction is shown for comparison.} \label{1D}
\vspace{-0.25in}
\end{figure}
The temperature evolution of the principal resonance line is shown in Figure~\ref{1D}, (a--c). At $T >T_N$, this mode narrows with the increasing temperature, concomitantly increasing in amplitude; its position does not change except in the vicinity of the ordering transition, at $T \lesssim 1.5T_N$. The integral intensity remains practically constant, in agreement with the low-$T$ susceptibility of a $S=1/2$ chain \cite{Motoyama_PRL1996,EggertAffleckTakahashi_PRL1994,Starykh_PRL1997}. Within the experimental accuracy, the linewidth appears to be independent of magnetic field (the excitation frequency), Fig.~\ref{1D}(b). Its temperature dependence is best described as $\mu_0\Delta H \simeq 0.014 +0.2/T$ (solid line). The $1/T$ contribution can be associated with a small anisotropy of the weak inter-chain coupling, $J^{\prime}_z \neq J^{\prime}_{x,y}$ \cite{FuruyaSato_JPSJ2015}, where our estimate yields $\delta J^{\prime}\sim 0.5$~K, while the constant term accounts for other contributions \cite{Supplementary}. According to Refs.~\cite{OshikawaAffleck_PRL1999,OshikawaAffleck_PRB2002,MaedaSakaiOshikawa_PRL2005}, the absence of a measurable $\sim 1/T^2$ contribution to the linewidth, as well as a very small observed $T-$dependent line shift, impose stringent upper limit on possible staggered fields, $h_{st}\lesssim 2\cdot 10^{-2}$~K, consistent with the ideal crystal structure of \SCO. The estimated $T-$linear contribution to the linewidth indicate an upper bound on the anisotropy of the intra-chain exchange, $\delta J/J \lesssim 1.4\cdot 10^{-2}$ \cite{OshikawaAffleck_PRB2002}, which is further confirmed by the analysis of the AFMR spectra observed below $T_N$ \cite{Supplementary}.

\begin{figure}[!t]
\centerline{\includegraphics[width=0.8\columnwidth]{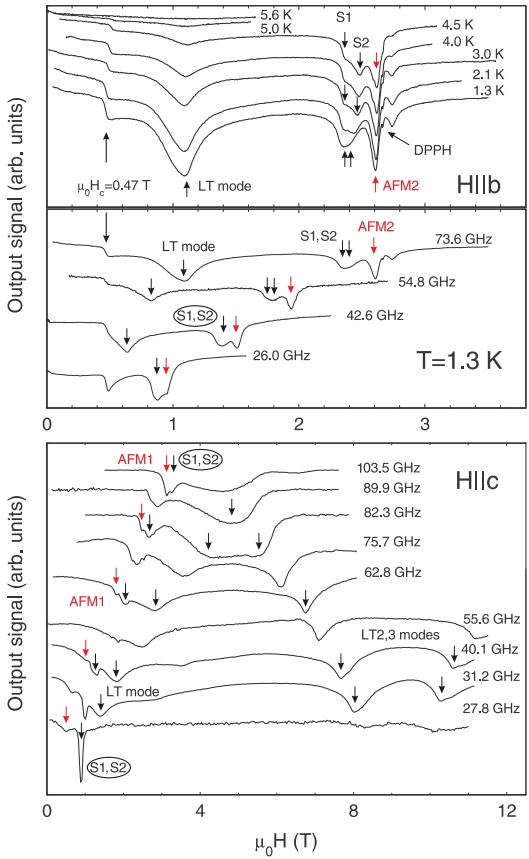}}
\caption{The resonance absorption spectra measured for $H\parallel b$-axis at frequency $\nu =73.6$~GHz on decreasing the temperature from above $T_N$ to 1.3~K (upper panel), and the low-temperature records made at various excitation frequencies for $H\parallel b$ (middle panel) and $H\parallel c$ (lower panel). Arrows mark absorption maxima corresponding to different resonance modes discussed in the text.
}
\label{spectra}
\vspace{-0.25in}
\end{figure}
At temperatures below the transition into an ordered state, the resonance spectrum of \SCO\ markedly transforms. We observe a gradual shift of the principal resonance line at all measurement frequencies, $\nu$, for $H\parallel a {\rm~and~} c$, and in the low-frequency range for $H\parallel b$, which is consistent with the opening of gaps in the spin excitations spectrum (see upper curves in Fig.~\ref{temp_c}) \cite{footnote1}. Assuming the relation for a field-dependent gapped mode in a two-sublattice antiferromagnet with weak anisotropy,
$\nu^2= \Delta^2 (T) + \left( g\mu_B/h \right )^2 H_{res}^2$, we obtain the temperature dependence of the gap, $\Delta (T)$, which is directly related to the AFM order parameter, $\Delta \propto\langle S\rangle$ \cite{ProzorovaBorovik_JETP1969}. The corresponding $\left( \Delta (T)/\nu \right)^2$ dependence for $H\parallel c$ and $\nu =27.0$~GHz is shown in Fig.~\ref{1D}(e), along with the intensity of the $(0, 0.5, 0.5)$ magnetic Bragg peak measured by neutron diffraction, which also probes the square of the magnetic order parameter. The excellent agreement between the two measurements confirms unambiguously that we observe the AFMR, and that magnon modes develop at $T < T_N$ in a system of weakly ordered chains in \SCO\ in the frequency range probed in our experiments. The fit of the square of the order parameter in the vicinity of the ordering transition reveals linear temperature dependence consistent with the chain mean field (CMF) theory \cite{Schulz_PRL1996}, and yields the N\'{e}el temperature $T_N = 5.5(1)$~K in agreement with previous studies \cite{Kojima_PRL1997}. The evolution of satellite modes S1, S2 below $T_N$ supports the conclusion that these signals originate from tiny amounts of defects and inclusions.

\begin{figure}[!t]
\centerline{\includegraphics[width=0.8\columnwidth]{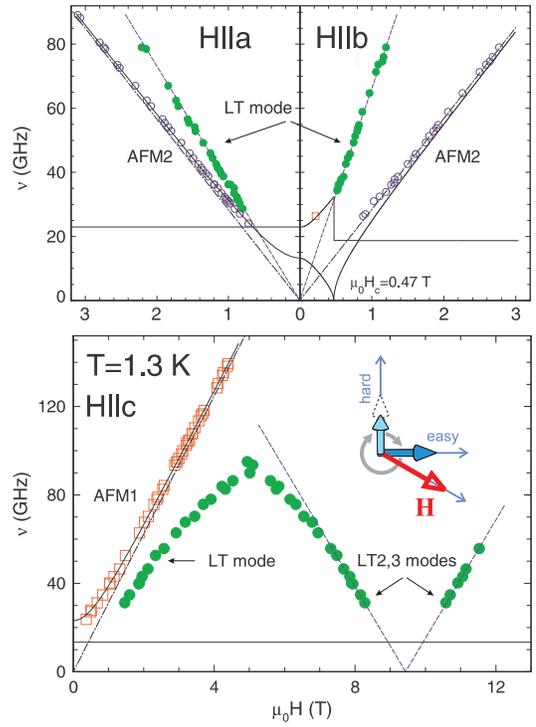}}
\caption{Frequency-field diagrams of the magnetic resonance spectra measured at $T=1.3$~K for the three principal directions of the applied field with respect to the crystal axes, $H\parallel a,b$ (top), and $H\parallel c$ (bottom). The two AFMR modes are shown with the open symbols, the solid lines are the theoretical calculations \cite{Nagamiya_1955,Andreev_JETP1978,AndreevMarchenko_1980} for the biaxial collinear antiferromagnet. Closed circles show the low-temperature LT mode; the dashed lines are linear fits discussed in the text. The dashed-dotted lines show the paramagnetic resonance position for the $g$-factors obtained from high-temperature measurements. The drawing in the lower panel illustrates a mechanism of dynamical coupling of the Higgs amplitude mode and the transverse Goldstone magnon corresponding to oscillations of the order parameter in the plane perpendicular to the applied field.}
\label{ffd}
\vspace{-0.25in}
\end{figure}
For $H\parallel b$, a step-like, non-resonant feature develops below $T_N$, in addition to the resonance modes (Fig.~\ref{spectra}, upper and middle panels). The magnetic field at which this feature arises, $\mu_0H_c\simeq 0.47$~T, does not depend on temperature and frequency. This allows its identification with a spin-flop transition, where a jump in the real part of magnetic susceptibility leads to a step-like absorption feature (note that at zero field the ordered magnetic moments in \SCO\ are directed along the $b$-axis \cite{Kojima_PRL1997}).

The behavior of the AFMR modes identified in our experiments for all directions of the applied magnetic field is equivalent to the spectrum of a collinear antiferromagnet with two inequivalent
anisotropy axes 
(stipulated by the orthorhombic symmetry of \SCO). The most general description of the low-energy spin dynamics of such a system is provided by the theory of spin hydrodynamics \cite{AndreevMarchenko_1980}. For a collinear antiferromagnet, such as \SCO\ at $T < T_N$, this phenomenological theory predicts two pseudo Goldstone transverse spin wave branches, which correspond to our AFMR modes 1 and 2. Their resonance excitation energies for all three directions of magnetic field are accurately reproduced by the theory using only two parameters, $\Delta_1=23.0$~GHz and $\Delta_2=13.3$~GHz, for the gap values rendered by a small anisotropy to the two Goldstone magnons (lines 1 and 2, respectively).
The critical field of the spin-flop transition, which is given by the relation, $\mu_0 H_c=h\Delta_2/(g^b\mu_B)\simeq 0.47$~T ($g^b$ is a $b$-component of the $g$-tensor), is in excellent agreement with the observed value.

Having thus unambiguously established the spinon confinement into the Goldstone magnons, giving rise to the AFMR in \SCO\ in the energy range covered in our measurements, $h\nu \lesssim k_BT_N$, we now focus on another remarkable feature. For all directions of the magnetic field, we observe an intense line of resonance absorption, marked ``LT mode'' in Fig.~\ref{spectra}. It has a roughly temperature-independent line width, several times larger than that of the AFMR modes. This line appears below $T_N$ and rapidly grows in intensity with no shift of the resonance field on further cooling. It reveals a novel magnetic excitation emerging in the ordered phase, which is both theoretically unanticipated and hitherto unobserved. For $H\parallel a,b$ this type of signal consists of a single resonance line; for $H\parallel b$ it only appears at $H>H_c$ (see middle panel of Fig.~\ref{spectra}). Two additional signals are observable in the high-field range for $H\parallel c$, whose intensity drops precipitously at $\nu \lesssim 30$~GHz (see data at 27.8~GHz and 31.2~GHz in Fig.~\ref{spectra}).

The frequency-magnetic field diagram of Fig.~\ref{ffd} shows the experimental data for the field dependence of the novel LT mode (closed symbols) along with that of the conventional AFMR lines 1, 2. For $H\parallel a,b$, the LT mode follows linear dependence, $h\nu =g_{eff}\mu_B H$, shown by dashed lines in the corresponding panels of Fig.~\ref{ffd}. The fit yields a very large and anisotropic ``effective $g-$factors'', $g_{eff}^a=2.60(5)$ and $g_{eff}^b=4.7(1)$. The triple line observed for $H\parallel c$ has a non-monotone field dependence, and appears to soften in high fields, possibly indicating a quantum phase transition. In the vicinity of the transition, $\mu_0H_{c2}\simeq 9.44$~T, the mode can be described by a critical-type linear dependence, $h\nu = g_{eff}^c\mu_B |H-H_{c2}|$, with a slope $g_{eff}^c=1.91(5)$ (dashed line in Fig.~\ref{ffd}).

The nature of the spin system in \SCO\ suggests that the new mode could be related to the amplitude fluctuations of the order parameter (Higgs mode) in a weakly ordered system of spin-1/2 AFM chains. We note that no such mode was observed in CsNiCl$_3$, a system of weakly coupled AFM spin-1 chains with a very similar $T_N \approx 4.8$~K, nor in a spin-5/2 chain system CsMnBr$_3$ ($T_N \approx 8$~K), where the ESR spectra are in full agreement with the spin hydrodynamics theory \cite{Zaliznyak_JETPLett1988,Zaliznyak_JETP1990}. In \SCO\ the theory predicts an undetectable (for our field-scanning technique), field-independent AFMR modes (horizontal lines in Fig.~\ref{ffd}), in addition to the observed field-dependent (pseudo Larmor) lines 1 or 2. An interaction of these Goldstone magnons with the amplitude (Higgs) mode provides a plausible mechanism by which they can acquire field dependent mass. At the origin of such coupling could be the spin anisotropy, which favors different amplitude of the ordered moment depending on its alignment with respect to the easy/hard axis. This is illustrated in the lower panel of Fig.~\ref{ffd}, where the Larmor precession of the ordered magnetic moment around the applied field modulates its amplitude.


In summary, the observed ESR response in the weakly coupled spin-1/2 chain antiferromagnet \SCO\ shows nearly ideal 1D Heisenberg behavior in the Luttinger-liquid phase. In the weakly ordered AFM phase at $T < T_N$, it reveals a novel, dominant excitation mode, which develops along with field-dependent gapped AFMR modes (transverse pseudo-Goldstone magnons) intrinsic to a collinear antiferromagnet with weak two-axial anisotropy. Recent analysis \cite{TsvelikZaliznyak_arXiv2016} of the Heisenberg necklace model performed in the context of the present findings suggests that such a strongly field-dependent excitation cannot be explained by a coupling to the paramagnetic impurities, or the nuclear spins, and therefore must be a property of the bulk spins. This new mode is missed by the macroscopic theory of spin hydrodynamics, which describes pseudo Goldstone spin waves. It thus embodies a short-wavelength physics, and can be understood as a mixed mode of the longitudinal and transverse (LT) fluctuations of the order parameter, resulting from the interaction of the Goldstone magnon with the Higgs amplitude mode. Consistent with this mixed character is its substantial width, reflecting the universally damped Higgs mode, and a large width of the longitudinal mode indicated by other experiments \cite{Lake_NatMater2005,Lake_PRB2005,Zheludev_PRL2002}. The observed softening of the novel LT mode at a critical field $H_{c2}$ might then herald a symmetry breaking transition to the longitudinal spin density wave state, which is expected in a system of weakly coupled spin-1/2 chains in a field \cite{Grenier_PRB2015}. While challenging the existing theories with disruptive novel results, 
our unexpected findings present exciting new opportunities for studying and understanding the interrelation of confinement, spontaneous symmetry breaking, and the Higgs physics.

\begin{acknowledgments}
The authors thank A.~I.~Smirnov, L.~E.~Svistov, V. N. Glazkov, M.~E.~Zhitomirsky, A. Abanov and A. Tsvelik for useful discussions. The work at P.~Kapitza Institute was supported by the Russian Fund for Basic Research,  Grant 15-02-05918, and the Program of Russian Scientific Schools. The work at Brookhaven National Laboratory was supported by the Office of Basic Energy Sciences, U.S.\ Department of Energy, under Contract No.\ DE-SC00112704. We acknowledge the support of NIST, US Department of Commerce, in providing the neutron research facilities used in this work.
\end{acknowledgments}

\bibliographystyle{apsrev4-1}

%

\end{document}